\documentstyle{article}
\begin{document}
\setlength{\baselineskip}{15pt}
\title{A constant of motion in 3D implies a local generalized Hamiltonian 
structure}
\author{Benito Hern\'{a}ndez--Bermejo \and V\'{\i}ctor Fair\'{e}n $^1$}
\date{}

\maketitle

{\em Departamento de F\'{\i}sica Fundamental, Universidad Nacional de 
Educaci\'{o}n a Distancia. Senda del Rey S/N, 28040 Madrid, Spain.}

\mbox{}

\begin{center} 
{\bf Abstract}
\end{center}

We demonstrate that a Poisson structure can always be associated to a general 
nonautonomous 3D vector field of ODEs by means of a diffeomorphism that 
preserves both the orientation and the volume of phase-space. The only 
prerequisite is the existence of one constant of motion. 

\mbox{}

\mbox{}

\mbox{}

{\bf Keywords:} Ordinary differential equations, generalized Hamiltonian 
structures, Poisson systems. 

\mbox{}

\mbox{}

\mbox{}

\mbox{}

\mbox{}

\mbox{}

\mbox{}

\mbox{}

\mbox{}

\mbox{}

\mbox{}

$^1$ To whom all correspondence should be addressed. E-mail vfairen@uned.es

\pagebreak
\begin{flushleft}
{\bf 1. Introduction}
\end{flushleft}

The possibility of associating a generalized Hamiltonian structure 
\cite{olv1,wei1} (also known in the literature as a Poisson structure) to 
general flows of ODEs has deserved a considerable attention recently, both 
from the point of view of the existence of such representations \cite{perl} 
and of their explicit determination in two \cite{cyf1}, three 
\cite{gyn1,hyg1,gyh1} and $n$ dimensions \cite{luc1,hoj1,hoj2}. The main 
advantage of Poisson structures lies in that they can encompass much wider 
categories of flows than the classical symplectic structure (of which the 
Poisson structure is just a generalization). However, this is done while 
preserving the benefits of the classical symplectic representation, as proved 
by Darboux' theorem \cite{olv1}.

The construction of generalized Hamiltonian formulations is of interest in  
the field of statistical mechanics, where it has been demonstrated that Nambu 
mechanics \cite{nam1} is just a particular case of Poisson structure 
\cite{ryk1,rug1,kob1}; in the Inverse Problem of Mechanics, with results 
which range from the classical Lie-Koenigs theorem for symplectic structures 
\cite{whi1} to the relationship between Poisson structures and Birkhoff 
systems \cite{san1}; in the discussion of nonlinear stability of relative 
equilibria by the energy-Casimir method \cite{hyw1,arn1,hym1}; in 
evolutionary dynamics, in which the connection between Poisson structures and 
replicator equations for bimatrix games has been recently established by 
Hofbauer \cite{hof1}; in quantization \cite{chi1}, due to the relevance of 
the link between Poisson and Jacobi manifolds; finally, may we mention the 
importance in numerical analysis where, in addition to the well-known 
symplectic integrators \cite{yos1,ssl1}, a new generation of algorithms 
is being developed \cite{qis1,qis2}, the efficacy of which is ensured 
automatically if the system admits a generalized Hamiltonian description. 

The Poisson structure for a system of first-order equations
\begin{equation}
   \label{nfl}
   \dot{x}^i = v^i(x^1, \ldots ,x^n,t) \; , \;\:\; i = 1, \ldots ,n
\end{equation}
is based: first, on an antisymmetric matrix ${\cal J}$ whose components 
$J^{ij}$ are, in general, functions of the $x^i$ and satisfy the Jacobi 
identities 
\begin{equation}
     \label{jac}
     J^{li} \partial_l J^{jk} + J^{lj} \partial_l J^{ki} + 
     J^{lk} \partial_l J^{ij}  = 0 \; ,
\end{equation}
where $ \partial_l $ means $ \partial / \partial x^l$, indices $i,j,k$ run 
from 1 to $n$ and the summation convention over repeated 
indices is applied; and second, on a Hamiltonian $H$, which is usually taken 
to be a time-independent first integral of the system. Then the flow 
(\ref{nfl}) is said to be Hamiltonian in the generalized sense if it can be 
written as
\begin{equation}
    \label{nham}
    \dot{x}^i = J^{ij} \partial _j H \; , \;\:\; i = 1, \ldots , n.
\end{equation} 
In other words, $v^i$ can be expressed as $v^i = [x^i,H]$ for the Poisson 
bracket defined by
\begin{equation}
   \label{bra}
   [F,G] \equiv ( \partial _i F) J^{ij} ( \partial _j G)
\end{equation}

Of course, the task of writting an arbitrary system (\ref{nfl}) in the form 
(\ref{nham}) cannot be accomplished in general, even if a first integral is 
known. This is also the case when the phase-space is three-dimensional 
(although this is the single dimension in which the Jacobi identities 
(\ref{jac}) are not overdetermined, as pointed out by Haas and Goedert 
\cite{hyg1}). 

We shall focus this work on the three-dimensional problem. In this case, it 
has been demonstrated that the knowledge of two independent first integrals 
is always sufficient to develop a generalized Hamiltonian formulation of the 
problem \cite{ryk1}. When only one first integral is known, the present-day 
methods merely allow the solution of the problem in certain situations 
\cite{gyn1,hyg1,gyh1}. We shall prove here the following general result: the 
knowledge of a {\em single\/} constant of motion leads locally 
to a Poisson structure. Given a first integral of the flow (\ref{nfl}), the 
usual approach considers a direct recasting of the system into Poisson form, 
thus limiting the scope of this format to just those few cases in which the 
procedure leading from (\ref{nfl}) to (\ref{nham}) can consistently be 
completed. Our solution to the problem does not intend to build the Poisson 
structure directly on the original system (\ref{nfl}) itself, as the previous 
procedures try to do. We instead overcome foregoing limitations by showing 
how an equivalent local Poisson structure can {\em always\/} be associated to 
an arbitrary three-dimensional system, if we first proceed to submit it to a 
simple orientation-preserving diffeomorphism, after which the hamiltonization 
procedure is straightforward. This implies automatically the strict 
topologically orbital equivalence between both flows \cite{jac}. Moreover, 
the diffeomorphism is also volume-preserving. The outcome is that it is 
always possible in practice to exploit the advantages derived from the 
Hamiltonian formulation of the problem, since both systems are dynamically 
equivalent and information can be directly transferred between them due to 
the invertibitily of the transformation.

\mbox{}

\begin{flushleft}
{\bf 2. Poisson structures in 3D flows}
\end{flushleft}

Due to its ubiquity in all fields of Physics, Chemistry and Biology, 3D 
systems have received an especial attention in the literature ---even an 
attempt of giving a sample of references about the issue would probably 
exceed the limits of this work. This has also been the case with regard to 
Poisson structures. One of the first fundamental results \cite{ryk1} showed 
that, in general, a 3D system possesses a Poisson structure if two 
independent first integrals exist. In a subsequent article, G\"{u}mral and 
Nutku \cite{gyn1} reduced the problem to the solution of a nonlinear partial 
differential equation, which can be transformed into a Riccati equation if 
two independent constants of motion are known. Later, Haas and Goedert 
\cite{hyg1} were able to reformulate the problem as the search of particular 
solutions of a {\em linear\/} partial differential equation. In their method 
(which will be our starting point) only one time-independent first integral 
of the system is, in principle, required; however, when two independent 
constants of motion are known the problem is reduced to quadrature. This 
approach was subsequently generalized by Goedert {\em et al.\/} \cite{gyh1} 
to include the case of time-dependent, rescalable first integrals. To make 
the list complete, we must also add those approximations which proceed 
directly from {\em ansatzs\/} for matrix ${\cal J}$, for the Hamiltonian or 
for both, such as that of Plank \cite{pla1}. 

The result of these efforts is that it has been possible to construct a 
Hamiltonian formulation for some 3D systems (or, at least, for certain 
integrable or semi-integrable cases of them). Apart from classical examples 
such as Euler's equations for a free rigid body \cite{arn1,sym1} or a 
particle in a magnetic field \cite{hoj1}, we may mention more involved (and 
interesting) systems like Lotka-Volterra equations \cite{pla1,pla2,nut1}, 
the Halphen system \cite{gyn1}, Maxwell-Bloch equations \cite{gyn1}, the 
Lorenz model \cite{gyn1,gyh1}, the Rabinovich system \cite{gyh1}, the RTW 
interaction \cite{gyh1}, and also some systems of biological relevance such 
as the Kermac-McKendric model for epidemics \cite{nut2} or the May-Leonard 
equations \cite{gyn1}. 

Consequently, any generalization of the previously mentioned techniques 
leading to the construction of Poisson structures for wider sets of flows 
would lead to an increase of the tools available for their analysis, as 
follows from the argumentation given in the Introduction. Such a 
generalization shall be our goal in the forthcoming sections of the paper. 
As mentioned before, our starting point will be the technique due to Haas 
and Goedert, which we briefly recall now for the sake of completeness. 

Let us consider a system of ODEs
\begin{equation}
     \label{ham2}
     \dot{x}^i = v^i(x^1,x^2,x^3,t) \;,\;\,\;\,\;\, i = 1, 2, 3 
     \;\; ,
\end{equation}
for which a time-independent constant of motion $H(x^1,x^2,x^3)$ exists. 
In short, what Haas and Goedert have shown is that, once $H$ is known, a 
completion of the problem is equivalent to finding one particular solution of 
the following linear partial differential equation, which is essentially a 
restatement of the Jacobi identities (\ref{jac}):
\begin{equation}
    \label{pde}
    v^i \partial _i J = AJ + B
\end{equation}
where 
\begin{equation}
    \label{ayb}
    A = \partial _i v^i  - \frac{ \partial _3 v^i 
    \partial _i H }{ \partial _3 H}  \; , \;\: 
    B = \frac{ v^1 \partial _3 v^2 - v^2 \partial _3 v^1 }{ \partial _3 H }
\end{equation}
Once a particular value of $J$ is known, system (\ref{ham2}) complies to 
format (\ref{nham}) with Hamiltonian $H$ and structure matrix:
\begin{equation}
   \label{laj}
    J^{12} = J \; , \;\: 
    J^{13} = \frac{v^1 - J \partial _2 H}{\partial _3 H} \; , \;\:
    J^{23} = \frac{v^2 + J \partial _1 H}{\partial _3 H} 
\end{equation}
The explicit determination of a solution of (\ref{pde}) for an arbitrary 
vector field is, in principle, not guaranteed. However, as Haas and Goedert 
point out \cite{hyg1}, equation (\ref{pde}) always admits the direct solution 
$J=0$ in the homogeneous case $B=0$, though they did not exploit this line of 
action. We shall do it here. This leads to a nontrivial solution of the 
Jacobi equations (\ref{jac}), since $J^{13}$ and $J^{23}$ do not vanish in 
general. Further details, such as those related to the preservation of scale 
invariance by equations (\ref{pde}) and (\ref{ayb}) can be found in the 
original reference \cite{hyg1}. 

\mbox{}

\begin{flushleft}
{\bf 3. Construction of a Poisson structure for 3D flows}
\end{flushleft}

We start by considering a system of the form (\ref{ham2}) for which a 
time-indepen\-dent $C^1$ constant of motion $I(x^1,x^2,x^3)$ is known:  
\begin{equation}
   \frac{\mbox{d}I}{\mbox{d}t} = v^i \partial _i I = 0
\end{equation}
Thus we can write $I(x^1,x^2,x^3)=I_0$, where $I_0=I(x^1(0),x^2(0),x^3(0))$. 
We shall assume without loss of generality that $\partial _3 I \neq 0$ (if 
this is not the case, a different labeling of the variables can always be 
adopted in order to exchange $x^3$ with either $x^1$ or $x^2$, as shown by 
Haas and Goedert \cite{hyg1}). 

As emphasized at the end of Section 2, the fundamental equation (\ref{pde}) 
does always admit a solution in the case $B=0$ or:
\begin{equation}
   \label{omm}
   v^1 \partial _3 v^2 - v^2 \partial _3 v^1 = 0
\end{equation}
This identity holds if, in particular:
\begin{equation}
   \label{noz}
   \partial _3 v^1 = \partial _3 v^2 = 0
\end{equation}
or, in other words, $v^i \equiv v^i(x^1,x^2,t)$ for $i=1,2$. This suggests 
the possibility of performing a change of dependent variables in system 
(\ref{ham2}) such that (\ref{noz}) holds for the target system. This 
transformation is:
\begin{eqnarray}
    \tilde{x}^1 & = & x^1                                 \nonumber    \\
    \tilde{x}^2 & = & x^2                                 \label{tr1}  \\
    \tilde{x}^3 & = & I(x^1,x^2,x^3) + \varphi (x^1,x^2)  \nonumber 
\end{eqnarray}
where $\varphi (x^1,x^2)$ is an arbitrary $C^1$ function. The condition 
$\partial _3 I \neq 0$ has two relevant implications. The first is that there 
exists a unique $C^1$ function $g$ such that the equation 
$I(x^1,x^2,x^3)=I_0$ can be equivalently written as: 
\begin{equation}
   \label{g}
   x^3-g(x^1,x^2,I_0) = 0
\end{equation}
This allows the explicit inversion of the third equation in (\ref{tr1}):
\begin{equation}
   \label{x3}
   x^3 = g(\tilde{x}^1,\tilde{x}^2,\tilde{x}^3- \varphi (\tilde{x}^1, 
   \tilde{x}^2))
\end{equation}
The second implication is that (\ref{tr1}) is one-to-one since the Jacobian 
is not zero. Consequently, transformation (\ref{tr1}) is a diffeomorphism. 

The equations of motion for the transformed system are then:
\begin{eqnarray}
   \dot{\tilde{x}}^1 & = & v^1(\tilde{x}^1,\tilde{x}^2,g(\tilde{x}^1,\tilde{x}^2,
      \tilde{x}^3 - \varphi(\tilde{x}^1,\tilde{x}^2)),t) \nonumber \\
   \dot{\tilde{x}}^2 & = & v^2(\tilde{x}^1,\tilde{x}^2,g(\tilde{x}^1,\tilde{x}^2,
      \tilde{x}^3 - \varphi(\tilde{x}^1,\tilde{x}^2)),t) \label{tar1} \\
   \dot{\tilde{x}}^3 & = & \dot{\tilde{x}}^1 \tilde{\partial} _1 
      \varphi(\tilde{x}^1,\tilde{x}^2) + \dot{\tilde{x}}^2 \tilde{\partial} _2 
      \varphi(\tilde{x}^1,\tilde{x}^2)  \nonumber
\end{eqnarray}
where $\tilde{\partial} _i$ denotes $\partial / \partial \tilde{x}^i$. It is 
straightforward to check that (\ref{tar1}) has the first integral:
\begin{equation}
   \label{ip1}
   \tilde{x}^3 - \varphi(\tilde{x}^1,\tilde{x}^2) = I_0
\end{equation}
This is, in fact, nothing else that the original constant of motion $I$ in 
terms of the new variables, as is evident from the third equation in 
(\ref{tr1}). Then, we can re-express (\ref{tar1}) in its final form:
\begin{eqnarray}
   \dot{\tilde{x}}^1 & = & v^1(\tilde{x}^1,\tilde{x}^2,g(\tilde{x}^1,\tilde{x}^2,
      I_0),t) \nonumber \\
   \dot{\tilde{x}}^2 & = & v^2(\tilde{x}^1,\tilde{x}^2,g(\tilde{x}^1,\tilde{x}^2,
      I_0),t) \label{tar2} \\
   \dot{\tilde{x}}^3 & = & \dot{\tilde{x}}^1 \tilde{\partial} _1 
      \varphi(\tilde{x}^1,\tilde{x}^2) + \dot{\tilde{x}}^2 \tilde{\partial} _2 
      \varphi(\tilde{x}^1,\tilde{x}^2)    \nonumber
\end{eqnarray}
This flow verifies condition (\ref{noz}) and it is then a Poisson system. 
From (\ref{ip1}), the Hamiltonian is $H = \tilde{x}^3 - 
\varphi(\tilde{x}^1,\tilde{x}^2)$. Notice how the arbitrariness of 
$ \varphi $ remains as an extra degree of freedom from which we can 
eventually profit to write the Hamiltonian in some desirable form. On the 
other hand, equations (\ref{laj}) provide the structure matrix:
\begin{equation}
\label{pemf}
   {\cal J} = \left( \begin{array}{ccc}
   0  &  0  &  \tilde{v}^1(\tilde{x}^1,\tilde{x}^2,t) \\
   0  &  0  &  \tilde{v}^2(\tilde{x}^1,\tilde{x}^2,t) \\
   - \tilde{v}^1(\tilde{x}^1,\tilde{x}^2,t) &
   - \tilde{v}^2(\tilde{x}^1,\tilde{x}^2,t) & 0
   \end{array} \right)
\end{equation}
where $\tilde{v}^i(\tilde{x}^1,\tilde{x}^2,t) = 
v^i(\tilde{x}^1,\tilde{x}^2,g(\tilde{x}^1,\tilde{x}^2,I_0),t)$ for $i=1,2$.
The reduction to a generalized Hamiltonian formulation is thus achieved.
Transformation (\ref{tr1}) is actually global, not dependent on any
particular value of the first integral $I$. On the contrary, the resulting
equations of motion (\ref{tar2}), and by extension the Poisson structure
matrix (\ref{pemf}), are particularized to the level surfaces of $I$, and
are thus parametrized by $I_0$. Consequently, in order to get (\ref{tar2})
and (\ref{pemf}), (\ref{tr1}) is to be applied locally, in each one of the
level surfaces.

Taking into account (\ref{g}), we may choose, in particular, the following 
representation for the constant of motion:
\begin{equation}
   \label{resi}
   I(x^1,x^2,x^3) = x^3 - g(x^1,x^2,I_0) + I_0 
\end{equation}
Once substituted into (\ref{tr1}), the Jacobian of the transformation takes 
the value $1$, and it is thus volume-preserving. Since this value is 
positive, both the original and the target system are also topologically 
orbital equivalent \cite{jac}. 

\mbox{}

\begin{flushleft}
{\bf 4. Additional considerations}
\end{flushleft}

We conclude our exposition by detailing some significant particular 
situations not considered in the previous section.

\mbox{}

{\em (I) Direct substitution of $x^3$ as a limit case}

Let us consider transformation (\ref{tr1}) in the case in which $I$ is 
written in the form (\ref{resi}). We can make the following choice for 
function $\varphi$:
\begin{equation}
   \label{vpi}
   \varphi (x^1,x^2) = g(x^1,x^2,I_0) - I_0
\end{equation}
Then the resulting system (\ref{tar2}) takes the form:
\begin{eqnarray}
   \dot{\tilde{x}}^1 & = & v^1(\tilde{x}^1,\tilde{x}^2,g(\tilde{x}^1,\tilde{x}^2,
      I_0),t) \nonumber \\
   \dot{\tilde{x}}^2 & = & v^2(\tilde{x}^1,\tilde{x}^2,g(\tilde{x}^1,\tilde{x}^2,
      I_0),t) \label{tar3} \\
   \dot{\tilde{x}}^3 & = & \dot{\tilde{x}}^1 \tilde{\partial} _1 
      g(\tilde{x}^1,\tilde{x}^2,I_0) + \dot{\tilde{x}}^2 \tilde{\partial} _2 
      g(\tilde{x}^1,\tilde{x}^2,I_0)         \nonumber
\end{eqnarray}
and the Hamiltonian is $H = \tilde{x}^3 - g(\tilde{x}^1,\tilde{x}^2,I_0) + 
I_0$. Note that this system is formally the same that would result when using 
equation (\ref{g}) to substitute directly $x^3$ in the original equations 
(\ref{ham2}). However, no change of variables is performed when we proceed in 
this way. We could say, alternatively, that the final variables are just the 
same than the original ones. The consistency of our general approach based 
on a change of dependent variables can then be checked easily since equations 
(\ref{tr1}) reduce, as expected, to the identical transformation when we 
substitute (\ref{resi}) and (\ref{vpi}) in them.

\mbox{}

{\em (II) 2D systems without a known first integral}

It is a well-known result that any 2D flow
\begin{equation}
     \label{2d}
     \dot{x}^i = v^i(x^1,x^2,t) \;,\;\,\;\,\;\, i = 1, 2
     \;\; ,
\end{equation}
possessing a time-independent first integral $I(x^1,x^2)$ can be written as 
a Poisson system \cite{cyf1,hyg1}. The analysis of Section 3 allows us to 
carry out this task when no first integral is known for equations (\ref{2d}) 
by means of a one-dimensional embedding, i.e., by recasting the flow 
(\ref{2d}) into a three-dimensional one. The standard recipe for this 
purpose consists of the addition of a new variable $x^3$ to the system. We 
do this in the following way:
\begin{eqnarray}
 & & \dot{x}^i = v^i(x^1,x^2,t) \;,\;\,\;\,\;\, i = 1, 2 \nonumber \\
 & & \dot{x}^3 = v^1 \partial _1 \varphi (x^1,x^2) + v^2 \partial _2 
     \varphi (x^1,x^2)  \label{23h}
\end{eqnarray}
where $\varphi (x^1,x^2)$ is again an arbitrary $C^1$ function. System 
(\ref{23h}) is analogous to (\ref{tar2}), and it is therefore Hamiltonian in 
the generalized sense with Hamiltonian function $H=x^3 - \varphi (x^1,x^2)$. 
On the other hand, (\ref{23h}) is obviously equivalent to the original flow 
(\ref{2d}) in each level surface $H =$ constant. 

\mbox{}

{\em (III) Quasipolynomial systems}

Let us consider 3D systems of the quasipolynomial form:
\begin{equation}
   \label{qp}
   \dot{x}^i = x^i \left( M_{i0}+\sum_{j=1}^{m} M_{ij} (x^1)^{B_{j1}} 
   (x^2)^{B_{j2}} (x^3)^{B_{j3}} \right) \;,\;\,\;\,\;\, i = 1, 2, 3 \;\: ,
\end{equation}
where $m$ is a positive integer and the rest of coefficients are assumed to 
be real. These systems have proven to be suitable for representing general 
nonlinear flows \cite{ker1,byv1}. We shall assume the existence of a first 
integral $I(x^1,x^2,x^3)$ verifying the condition $\partial _3 I \neq 0$. If 
we substitute equations (\ref{qp}) in (\ref{omm}) we find:
\[
\sum _{i,j=1}^m (M_{1i}M_{2j} - M_{1j}M_{2i})B_{j3} x_1^{B_{i1}+B_{j1}} 
   x_2^{B_{i2}+B_{j2}} x_3^{B_{i3}+B_{j3}-1} \; +
\]
\begin{equation}
  + \; \sum _{k=1}^m ( M_{10} M_{2k} - M_{20} M_{1k} ) B_{k3} 
   x_1^{B_{k1}} x_2^{B_{k2}}x_3^{B_{k3}-1} = 0
\end{equation}
This leads to the nonlinear system of equations:
\begin{equation}
  \begin{array}{ll}
      (M_{1i}M_{2j} - M_{1j}M_{2i})(B_{j3} - B_{i3}) = 0 \;\:, & 
            i,j = 1, \ldots , m \; ; \; \; i \neq j \\
      (M_{10} M_{2k} - M_{20} M_{1k})B_{k3} = 0 \;\:, & 
            k = 1, \ldots , m 
  \end{array}
  \label{sqo}
\end{equation}
There are two possibilities: 

{\bf i)} There is at least one value of $i$, $1 \leq i \leq m$, such that 
$B_{i3}\neq 0$. Then equations (\ref{sqo}) can be used to demonstrate that 
there exists a real constant $\xi$ such that (row2) $=$ $\xi$ $ \times$ 
(row1) in matrix $M$. From this it is straightforward to prove that a first 
integral of the form $(x^1)^{\xi}(x^2)^{-1}$ exists. Since this constant of 
motion verifies $\partial _3 [ (x^1)^{\xi}(x^2)^{-1}] = 0$, it must be 
different to $I$. Then system (\ref{qp}) has two independent constants of 
motion and it is integrable.

{\bf ii)} $B_{i3} = 0$ for all $i$. Then equations (\ref{sqo}) are always 
satisfied.

The only nontrivial situation is therefore the second one, which implies, to 
sum up, that $x^3$ is not present in the first two equations of system 
(\ref{qp}). This is precisely the meaning of equation (\ref{noz}). 
Consequently, conditions (\ref{omm}) and (\ref{noz}) are completely 
equivalent, up to trivial and nongeneric cases, for quasipolynomial systems 
of the general form (\ref{qp}). 
\pagebreak
\begin{flushleft}
{\bf 5. Final remarks}
\end{flushleft}

In the preceding sections we have demonstrated how, for a 3D system of ODEs 
possessing one constant of motion, but for which a generalized Hamiltonian 
representation may not exist, such a structure can be achieved when the 
original vector field is suitably transformed to a completely equivalent 
system, in such a way that the information obtained (both qualitative and 
quantitative) from the Hamiltonian formulation can be carried back into the 
original flow. This approach seems to be relatively new in the literature, 
since the usual techniques try to build the Hamiltonian structure for the 
original system itself. The result is that the problem can be solved only 
for a limited set of ODEs or, in the case of the most general approaches, 
that the practical implementation of the Poisson structure is exceedingly 
difficult. One exception to this trend is the work of Cair\'{o} and Feix 
\cite{cyf1}, where they are able to construct a symplectic structure for 2D 
flows by means of a rescaling of time. However, it is the authors' conjecture 
that additional transformations on variables apart from that on time are 
necessary in order to handle the problem in three and higher dimensions. In 
this sense, an appropriate manipulation of the phase-space variables is one 
of the most attractive (perhaps the unique) possibility.

\mbox{}

\mbox{}

\begin{flushleft}
{\bf Acknowledgements}
\end{flushleft}

This work has been supported by the DGICYT (Spain), under grant PB94-0390. 
B. H. acknowledges a doctoral fellowship from Comunidad Aut\'{o}noma de 
Madrid. The authors also acknowledge Drs L. Cair\'{o}, F. Haas, M. Plank and 
G. R. W. Quispel for supplying them with copies of their works.

\pagebreak

\end{document}